# TOWARDS SCALABLE EM-BASED ANOMALY DETECTION FOR EMBEDDED DEVICES THROUGH SYNTHETIC FINGERPRINTING

Kurt A. Vedros[1], Georgios Michail Makrakis[1], Constantinos Kolias[1], Robert C. Ivans[2], and Craig Rieger[2]

[1]Department of Computer Science, University of Idaho, 1776 Science Center Dr, Idaho Falls, ID 83402 USA
kvedros@uidaho.edu, gmakrakis@uidaho.edu, kolias@uidaho.edu
[2]National and Homeland Security, Idaho National Lab, 1955 N Fremont Ave, Idaho Falls, ID 83402 USA
robert.ivans@inl.gov, craig.rieger@inl.gov

## ABSTRACT

*Embedded devices are omnipresent in modern networks including the ones operating inside critical environments. However, due to their constrained nature, novel mechanisms are required to provide external, and non-intrusive anomaly detection. Among such approaches, one that has gained traction is based on the analysis of the electromagnetic (EM) signals that get emanated during a device's operation. However, one of the most neglected challenges of this approach is the requirement for manually gathering and fingerprinting the signals that correspond to each execution path of the software/firmware. Indeed, even simple programs are comprised of hundreds if not thousands of branches thus, making the fingerprinting stage an extremely time-consuming process that involves the manual labor of a human specialist. To address this issue, we propose a framework for generating synthetic EM signals directly from the machine code. The synthetic signals can be used to train a Machine Learning based (ML) system for anomaly detection. The main advantage of the proposed approach is that it completely removes the need for an elaborate and error-prone fingerprinting stage, thus, dramatically increasing the scalability of the corresponding protection mechanisms. The experimental evaluations indicate that our method provides high detection accuracy (above 90% AUC score) when employed for the detection of injection attacks. Moreover, the proposed methodology inflicts only a small penalty (-1.3%) in accuracy for the detection of the injection of as little as four malicious instructions when compared to the same methods if real signals were to be used.*

## KEYWORDS

*Side Channel Analysis, Anomaly Detection, Electromagnetic Signals, Synthetic Signals*

## 1. INTRODUCTION

Nowadays, a large portion of corporate, government, military and critical infrastructure networks consist of embedded devices. Typically, these mission-critical assets are severely constrained in terms of processing, memory, and energy resources. Since standard cryptographic algorithms were designed according to the hardware specification of high-end systems, traditional crypto libraries and the corresponding protection tools are not applicable (at least not without modifications). At the same time, in many cases, embedded devices are directly exposed to the Internet and its cyber threats. Therefore, there is a dire need for the development of novel security mechanisms specifically designed to respect the limitations and peculiarities of such critical systems.

As a potential solution to this problem, researchers have relied on the analysis of patterns of analog signals emitted by the CPU of embedded devices. In this context, such signals are considered a side-channel because they get emitted involuntarily by devices during their regular operation. Even though these analog signals are often treated as noise in most applications, they may bear valuable information. In principle, certain characteristics of the emitted analog signals have a strong correlation to the instructions being executed by the CPU. Thus, numerous side-channel based anomaly detection approaches have been proposed particularly to provide external protection for embedded devices [1], [5], [7], [8], [9], [13], [14], [16].

Today, the dominant methods of side-channel based anomaly detection rely on the analysis of power-consumption patterns. This is primarily due to the ease of data collection and the robustness of this modality against external noise. Nevertheless, when compared to the power-based approach, electromagnetic (EM) based methods are theoretically more advantageous because the EM spectrum offers higher bandwidth, and the EM signals can be sampled at higher rates [13][5]. Moreover, depending on the type of antenna, the approach can be less invasive as the monitoring can be performed from a distance in real time. In fact, EM-based anomaly detection tools have proven to be successful for the detection of |kk: change refs here| extensive [7], [8], or even minimal modifications, say, down to the injection of a few instructions (assembly level) [15], [12].

Nevertheless, the development of EM-based defenses and the deployment of corresponding real-life solutions remain stagnant due to the limitations of traditional workflows. More specifically, a well-known challenge of these approaches revolves around the requirement for exhaustive fingerprinting of all normal execution states of the targeted program. This issue is severely neglected by the research community although it may be one of the most important practical roadblocks that prevent corresponding tools from being deployed in real life. To address this issue, we introduce a novel framework for generating synthetic EM signals directly from machine code. Most importantly, the generated synthetic signals can be used instead of real ones for anomaly detection purposes as part of the model training/fingerprinting stages. In further detail, our approach relies upon first constructing a library of the EM signatures of minimum execution units (i.e., in this case, assembly instructions) that can be used to synthesize the EM footprint of longer sequences of code. The advantage of the proposed approach is that it completely removes the need for an elaborate and error-prone fingerprinting stage. The EM signals used for training do not need to be captured, but rather they are *inferred* directly from ASM code in an offline step. This fact alone makes the entire process extremely scalable.

In summary, the main contributions of this work are (a) the identification of the requirements and structure of a database of signal blocks that can be used for the generation of synthetic EM sequences, as well as (b) a methodology for properly synthesizing such sequences of instructions corresponding to entire execution paths/code-sequences.

## 2. PROBLEM STATEMENT AND THREAT MODEL

Typically, software supporting embedded devices designed to control critical processes is considered of *low complexity* when compared to the analogous software running on servers and desktop systems. Indeed, corresponding workflows involve cycles of sensing, processing, and then acting, all executed in a loop fashion. However, realistically, even the simplest examples of this family of software may be comprised of hundreds if not thousands of execution paths spawned by conditional branching instructions. When the objective is to fingerprint the characteristics of normalcy then all of these execution paths must be observed and modeled. Particularly, EM-based fingerprinting is mainly a human-expert centric process revolving around tasks such as the correct positioning of probes, deciding the optimal recording parameters like the sampling rate, and synchronizing EM signals, among others. This, in turn, renders EM fingerprinting an extremely time-consuming, error-prone, and costly process.

This challenge is further amplified by two real-life restrictions. Firstly, execution branches may exist in a program that by definition are meant to be rarely followed. Even techniques such as the forceful execution [6] of specified branches might not be an option as such may be associated with critical failures. For this reason, these branches are likely to be left out of the fingerprinting phase. In this case, the resulting models will yield wrong predictions for these normal yet rare situations. Secondly, embedded devices occasionally receive firmware/software updates. These modifications in the executable generate the requirement for fingerprinting the behavior of the device from scratch. These challenges are illustrated in Figure 1.

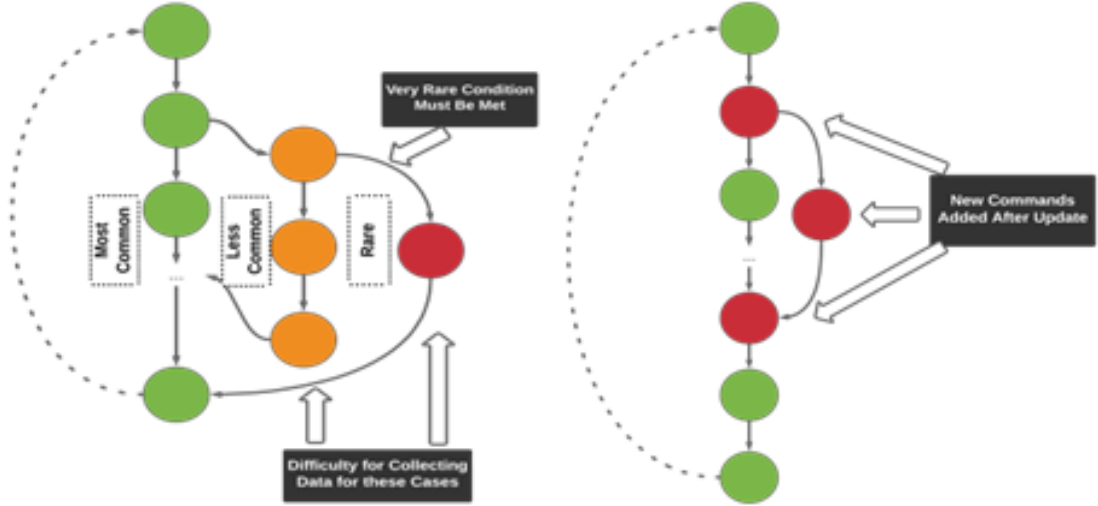

Figure 1. Scenarios where fingerprinting which is based on synthetic data could be valuable. Rare execution paths of programs are depicted in red and orange (left). New states/commands introduced after software updates are depicted in red (right).

As a solution, this work proposes a methodology for conducting the fingerprinting stage completely offline through the use of synthetically generated EM signals. More formally, our approach assumes that a mapping of instructions to signals $\mathcal{M}$ exists such that

$$\mathcal{M} = \begin{bmatrix} c_1 & S_1^{c_1} \\ c_2 & S_2^{c_2} \\ \dots & \dots \\ c_n & S_n^{c_n} \end{bmatrix}$$

where $c_i$ is an instruction supported by a specific processor architecture and $S^{c_i} = \{s_1, s_2, \dots, s_m\}$ is the corresponding EM signal observed when the instruction ci gets executed by the CPU. The signal gets described by a sequence of samples $s_i$ whose number depends on the sampling rate and the duration of the corresponding instruction (most architectures support instructions of variable duration).

The task at hand is to discover a function $f$ such that when given a sequence of instructions $I$ then it produces a version $S'$ that is similar to the real signal $S$ observed when $I$ is executed in the CPU. In other words, $f(I) = S'$ with the constraint that $D(S, S') \approx 0$, where $D$ is a distance metric.

Theoretically, the two main challenges of this approach are that (a) the number of instructions contained in $\mathcal{M}$ must be exhaustive. An additional challenge is that (b) a large number of signals corresponding to the same instruction $c_i$ must be captured because the phenotype of signals corresponding to the same sequence of instructions is not static.

In this work, we examine a specific application that can be supported by the proposed approach namely, *anomaly detection*. More specifically, we consider the situation where the attacker has discovered a vulnerability in the code that allows them to perform *code injection*. In practice, this is typically achieved by exploiting *buffer overflow* vulnerabilities. We assume that the attacker can inject an arbitrary number of instructions at any position of a target branch. Moreover, even a minimal number of instructions can have a meaningful malicious impact. We recognize, that in reality, while this situation is possible the attacker will usually have less flexibility.

Compared to the generalized version of the task, this problem has a more relaxed constraint i.e., assuming that the malicious version of the corresponding EM signal $S_m$ and an unmodified (normal) version $S_n$ then the following condition $D(S_m, S_n) > D(S_n, S'_n)$ is true. Regardless, the same concept can be applied to applications beyond anomaly detection such as side-channel analysis for inferring cryptographic keys.

## 3. PROPOSED FRAMEWORK

The purpose of the proposed framework is to conduct anomaly detection with high accuracy using synthetically generated versions of the EM signals that correspond to the *normal execution branches only*. A high-level overview of the proposed framework is given in Figure 2. In summary, the main steps involved in the process are as follows. During an offline step, a database of instructions-to-signal correlations is created (this is denoted as step ❶ in Figure 2). Next, synthetic signals are created using the database of EM signals and the target binary (step ❷ in Figure 2). Then, the synthetically generated sequences are used to train the baseline during the fingerprinting phase (step ❸). After this sequence, the target device is expected to be deployed on the field. At this point, the anomaly detection phase takes place (step ❹). At this point, real EM signals emanated by the device are captured once again, this time to be evaluated for anomalies. This process also capitalizes on the baseline that is already created during the previous step. Under the hood, the process involves the execution of machine learning algorithms that judge whether the new signal bears significant morphological similarities with the synthetic ones that were used to create the baseline. Hereunder, we shall analyze the basic steps of the process in further detail.

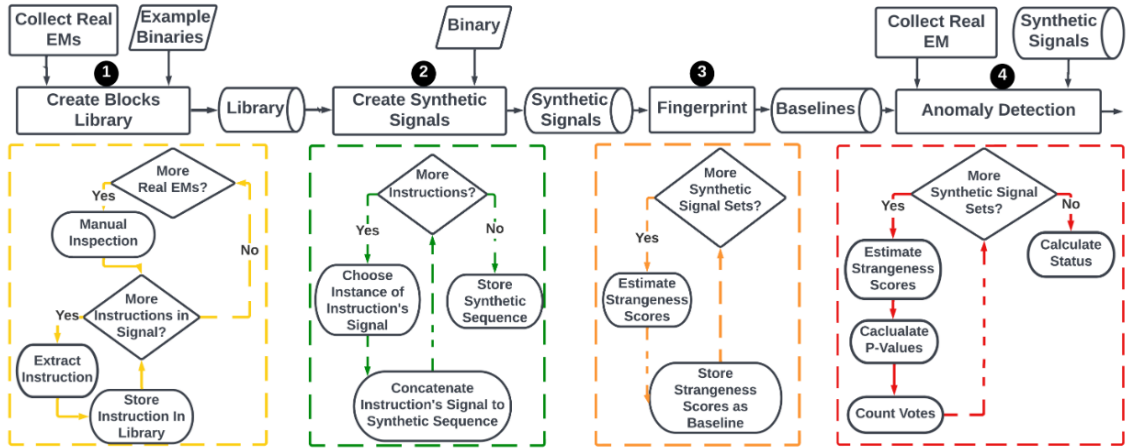

Figure 2. Workflow of the proposed framework.

The proposed framework assumes that a reliable mechanism for capturing EM signals from the elements of devices (e.g., CPU) is available. Today, this can be achieved by solely relying on Commercial-Of-the-Self (COTS) components. Such an assembly of components typically consists of (a) a near-field antenna for gathering the raw signals, (b) an amplifier for increasing the strength of the captured signal, (c) an oscilloscope for digitizing the collected analog signals, (d) and finally hard disks for storing the captured signals in their discrete form. In this framework, the process of capturing signals is performed in two separate stages i.e., during the construction of the library of basic building blocks (step ❶), a process which is completed offline, and during run-time for actively monitoring the health status of a target device (step ❹). Typically, the signals are collected by placing the antenna in close proximity to the CPU. However, in more complicated settings signals can be collected from multiple onboard components (e.g., the network module), and create more advanced correlations regarding the behavior of the device. Particularly for the latter case, an extra step of pre-processing that may involve noise elimination procedures may be included as part of steps ❸ and/or ❹ In this work, we have omitted such processes for purposes of simplicity.

### 3.1. Building a Library of Singal Blocks

A library of *basic building blocks* of signals is assumed to have been created in an offline step, a priori. This library should be available during the *fingerprinting* of any program, or more accurately any subsequence of any execution path inside a program. Theoretically, the term *basic building block* corresponds to the EM signature of each assembly level instruction, e.g., *and*, *nop*, etc.

Experimentally, we have identified that the main challenge with this approach is that one instruction in a sequence influences the shape and amplitude characteristics of the subsequent instructions. Typically, the directly subsequent instruction is influenced only. However, depending on the type of instruction (e.g., instructions involved in IO operations) multiple subsequent instructions may also be affected but to a lesser extent. In this work, we have assumed that only one instruction gets affected for reasons of simplicity, but further investigation is required. Therefore, the structure of the database introduced previously can be more accurately redefined as:

$$\mathcal{M} = \begin{bmatrix} (c_1 \mid c_1) & S_1^{(c_1|c_1)} \\ \cdots & \cdots \\ (c_2\|c_1) & S_1^{(c_2|c_1)} \\ \cdots & \cdots \\ (c_{n-1}\|c_n) & S_1^{(c_{n-1}|c_n)} \end{bmatrix}$$

where the $(c_{i-1} \mid c_i)$ operation indicates that instruction $c_i$ has been observed after

$c_{i-1}$. The reader should notice that an entry $S^{(c_{i-1}|c_i)}$ is comprised by the same number of samples as $S^{c_i}$.

Let us examine the requirements for the construction of such a database through a simple example. The x86 architecture supports 981 [4] unique instructions while a simpler CPU architecture like AVR includes unique 123 instructions [10]. Let us focus on the AVR architecture since it is widely deployed in embedded systems. Let us assume that 1000 examples of each instruction are captured then the original size of the database is estimated to have $1000 * 123 = 123K$ entries. In the lieu of this restriction, the database needs to have a total of $1,000 * 123^2 \approx 15M$ entries which is approximately two orders of magnitude larger than the original estimation. It is obvious, that the process of creating a database of all possible instructions is time-consuming. Regardless, this needs to be conducted only once. One can argue that once constructed, a database for a specific architecture can be open-sourced and made publicly available. Moreover, in practice, certain instructions are never observed together, while there are certain combinations of instructions that are observed much more often together. Thus, the requirements of constructing such a database are not prohibitive.

## 3.2. Generating Synthetic EM Signals

The process of generating synthetic signals for anomaly detection is as follows: (a) based on the sequence of instructions included in the binary, identify the next instruction that will be executed, (b) fetch a random EM sample that is associated with this instruction from the library, and (c) appended the EM at the end of a collective synthetic signal. The above steps are repeated until no more instructions are contained in the target sequence.

---

**Algorithm 1** Fingerprinting phase

---

1: **function** STRANGENESS:(benign dataset $X$, set of query signals $Q$, number of neighbors $\kappa$):
2:     $Strangeness\_Scores = []$
3:     for $\forall q \in Q$ do
4:         $D = []$
5:         for $\forall x \in X$ do
6:             $D \leftarrow distance(x, q)$
7:         end for
8:         $Nearest\_Neighbors = get\_min(D, \kappa)$
9:         $Strangeness\_Scores \leftarrow Sum(Nearest\_Neighbors)$
10:     end for
       return $Strangeness\_Scores$
11: **end function**
12: **function** FINGERPRINT:(benign dataset $X_s$, number of neighbors $\kappa$, number of benign execution paths $s$):
13:     $Baselines = []$
14:     for $\forall i \in s$ do
15:         $Baselines \leftarrow strangeness(X_i, X_i, \kappa)$
16:     end for
       return $Baselines$
17: **end function**

---

## 3.3. Fingerprinting Phase

The discovery of malicious EM signals was approached as a *semi-supervised anomaly detection* problem as opposed to *a supervised classification* problem. The reason for this decision is that nearly infinite alterations to a benign program can be performed by an attacker. This makes collecting instances of all possible known and unknown malicious versions of a program unrealistic. However, since the normal modes of operation of a device are finite, it is valid to assume that the corresponding EM signals can be collected, or in the context of this work, be synthetically generated. Therefore, we relied on and extended an existing semi-supervised anomaly detection method [2]. This method is based on the principles of *transduction* and *hypothesis testing*. Transduction is a technique of placing a sample in a set of known normal observations and understanding whether that sample is a good fit in the set. From the perspective of our experiment, the terms sample and observations refer to EM signals.

The method calculates a distribution of normalcy, namely, a baseline, between all the known benign cases corresponding to the same operational mode (i.e., an entire or parts of the same execution path). Realistically, a program can have several execution paths with each execution path corresponding to a different aspect of normal operation. This in turn, likely creates a unique EM signal.

In further detail, during this phase, a set of benign signals, X, is provided for each execution path. X must contain a significantly large number of EM signals because as explained in previous sections, recordings can deviate due to random phenomena occurring during the capture. In order to calculate the distribution, the *strangeness* (similarity) score of each sample point x with the rest in X must be calculated. Any algorithm that calculates the similarity (e.g., euclidean distance) can be used to estimate the *strangeness*. These include rudimentary approaches such as the mean of distances, or more sophisticated metrics like Local Outlier Factor [3] (which internally rely on euclidean distance). The processes involved in the fingerprinting phase are given in Algorithm 1. We relied on the sum of the $\kappa$ nearest (most similar) neighbors (signals) and utilized the euclidean distance metric. The outcome of this process is one (or multiple) lists that contain the similarity scores, referred to as *Strangeness_Scores*, (lines 5-9) in the algorithm. The *Strangeness_Scores* reflects the distribution of normalcy or simply put a baseline, (lines 13-16). This process is repeated for all possible execution paths.

---

**Algorithm 2** Anomaly Detection phase

---

1: **function** K-NN STROUD:(sets of benign signals $X_s$, strangeness baselines $B_s$, signal for evaluation $q$, number of neighbors $\kappa$, threshold $\tau$, number of benign execution paths $s$ ):
2:   $Votes = []$
3:   **for** $\forall i \in s$ **do**
4:      $size = length(B_i)$
5:      $score_q = strangeness(X_i, q, \kappa)$
6:      $Sorted\_Baseline_i = sort(B_i, ascending)$
7:      $index = 0$
8:      **for** $\forall score_x \in Sorted\_Baseline_i$ **do**
9:         **if** $score_q < score_x$ **then**
10:           $index = index + 1$
11:         **end if**
12:      **end for**
13:      $p\_value \leftarrow \frac{1 + size - index}{1 + size}$
14:      **if** $p\_value > \tau$ **then**
15:         $Vote \leftarrow normal$
16:      **else**
17:         $Vote \leftarrow anomalous$
18:      **end if**
19:      $Votes \leftarrow Vote$
20:   **end for**
21:   $status = anomalous$
22:   **for** $\forall vote \in Votes_q$ **do**
23:      **if** $vote = normal$ **then**
24:         $status = normal$
25:      **end if**
26:   **end for**
      **return** $status$
27: **end function**

---

### 3.3. Anomaly Detection Phase

The deployment phase assumes that the baselines, $B_s$, have already been produced successfully during the fingerprinting phase. Furthermore, the original sets of benign signals used to create the baselines, $X_s$, and the number of benign execution paths, $s$, are provided. Additionally, a signal for evaluation, $q$, is available. Finally, user-provided parameters that correspond to the number of neighbors ($\kappa$) and the threshold used to separate the normal from abnormal ($\tau$) signals are given. The overall process is provided in Algorithm 2.

During this process, a benign signal set of each execution path, $X_i$, is obtained from $X_s$. Then the strangeness of the new observation, *score $_q$* , is evaluated by comparing q to $X_i$ using the same algorithm implemented in the Fingerprinting phase, (line 5). Afterward, *score $_q$* is compared against the respective baseline, $B_i$, that was created by $X_i$ in the Fingerprinting phase. The comparison process is executed using transduction, creating a *p_value* for *q*, (lines 8-13). If the *p_value* is above the threshold $\tau$, then *q* is considered within the norm of the execution path, and a vote is saved as normal. Otherwise, the *vote* is saved as abnormal, (lines 14-19). This process is repeated for all execution paths in $X_s$ to see if *q* falls within the norm of *any* of the benign execution paths. Under normal conditions, benign signals are expected to be considered normal for one execution path. As such, only one *vote* for the unknown signal *q* being normal is required to flag it as benign. If no *vote* is given, then *q* is flagged as anomalous, (lines 21-26).

The voting mechanism was an extension to the original algorithm implemented to account for the certainty of a program being comprised of numerous paths. A comprehensive fingerprinting of a target program must consider as *normal* all possible paths inside that program.

## 4. EXPERIMENTAL EVALUATION

### 4.1. Testbed

The *target platform* used, was an Arduino Mega device. This device is equipped with an 8-bit ATmega2560 AVR microcontroller unit (MCU). This family of MCUs is a popular choice for both research as well as real-time control applications [11]. To acquire EM signals, we made use of an EMRSS RF Explorer H-Loop *EM probe*, which was placed exactly on top of the CPU. Since emanations from the CPU have a very low amplitude, each signal that we acquired was first amplified using a Beehive 150A EMC probe *amplifier* and then saved in a digital format using a PicoScope 3403D *oscilloscope* and a *laptop*. The chosen sampling rate was 500 MSamples/sec, the sampling interval is $2\ nsec$, and the average duration of the programs considered (loop portion only) was less than $1\ \mu sec$. Notice that we obtained the sample in a virtual noise free environment. In accordance with Vedros et al. [15] these sampling rates are able to detect the corruption of a program through the injection of even a single instruction in relatively noisy environments without the need for noise reduction pre-processing. The experimental setup can be seen in Figure 3.

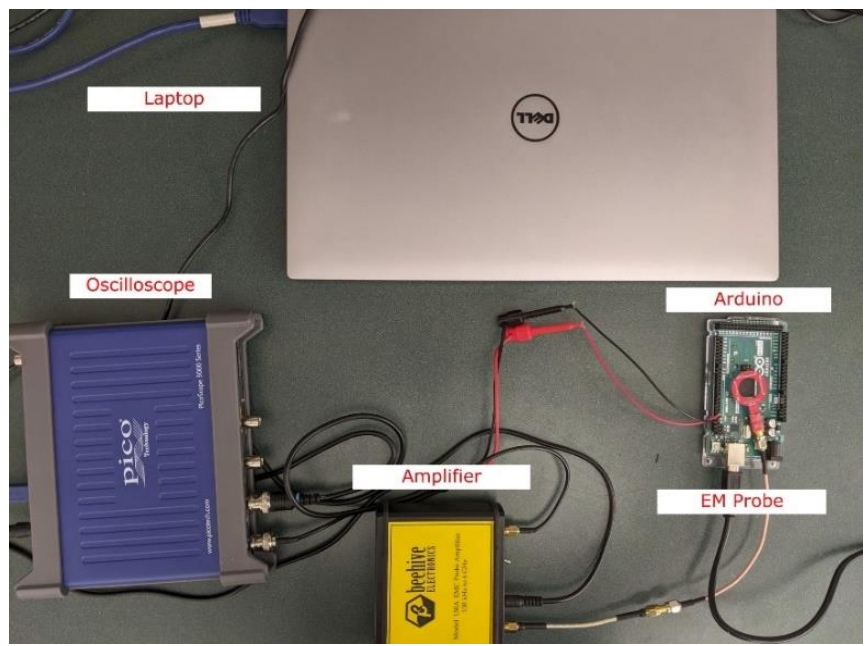

Figure 3. Experimental setup used for all the signals acquisitions described in this paper.

## 4.2. Test Cases

We considered the following scenario: a benign program with just one execution path is already installed in the target platform. The original software is comprised of just *17 instructions being executed inside a loop*. At some point, there was a need to modify the original program. In the update *several instructions were substituted, a new was added and one was removed from the original sequence*. The task is to synthetically generate EM signals of the modified version of the program directly from the ASM code so that we do not have to engage in the data-gathering process, from scratch. The original (*Program A*) and the updated version of the program (*Program B*) are given in Figure 4.

```
1  setup:
2    sbi  ddrb, 6  ; set pb6 as output (our
         sync artifact)
3  loop:
4    sbi   pinb, 6
5    ; simulates a decision
6    clr   r20
7    ldi   r20,  1
8    ; Will go to the first_label
9    ldi r22, 1
10   cp   r20, r22
11   breq  first_label
12   rjmp loop
13
14 first_label:
15   clr r2
16   ldi   r23, 1
17   mov  r1, r23
18   cp r1,  r2
19   lsl   r1 ; multiply r1 by 2 with logical
         shift left
20   lsr   r2 ; divide r2 by 2 with logical
         shift right
21   ses ; Set signed flag
22   cls ; Clear signed flag
23   sev ; Set Overflow Flag
24   clv ; Clear Overflow Flag
25
26   rjmp loop
```

```
1  setup:
2    sbi  ddrb, 6  ; set pb6 as output (our
         sync artifact)
3  loop:
4    sbi   pinb, 6
5    ; simulates a decision
6    clr   r20
7    ldi   r20,  1
8    ; Will go to the first_label
9    ldi r22, 1
10   cp   r20, r22
11   breq  first_label
12   rjmp loop
13
14 first_label:
15   ldi r23, 0
16   and   r2, r3 ; Bitwise AND (result in
         stored r2)
17   add  r1, r2 ; Add r2 to r1 (r1=r1+r2)
18   eor   r2, r3 ; Bitwise exclusive or
         between r2 and r3
19   sub  r1, r2 ; Subtract r2 from r1
20   ses ; Set signed flag
21   cls ; Clear signed flag
22   sev ; Set Overflow Flag
23   clv ; Clear Overflow Flag
24   clr r1
25
26   rjmp loop
```

Figure 4. The original version of the program (left) and the version of the program after the update (right). Different instructions are highlighted in red.

Next, we assumed that a malicious entity performs a modification (i.e., code injection) to our program. To illustrate the occurrence of such an attack, we developed two contaminated versions of the updated program *(Program B)*, each with differing amounts of injected code. The first contaminated version assumes that four malicious instructions were injected, while in the second case we have the injection of only two instructions. Consequently, the second version will be harder to detect due to the shorter length of the foreign code. The point of injection for both contaminated versions is in the middle of the sequence of the ASM instructions. The two malicious programs (*easy and hard*) are given in Figure 5.

Table 1. The number of samples and number of time indexes in each set.

| | Examples | Samples |
|---|---|---|
| Program A | 1000 | 1261 |
| Program B | 1000 | 1261 |
| Synthetic B | 1000 | 1261 |
| Malicious B Easy | 1000 | 1511 |
| Malicious B Hard | 1000 | 1386 |

```
1   ...
2   ; up to here, same as "Program B"
3
4   add  r1, r2 ; Add r2 to r1 (r1=r1+r2)
5   ;===== 4 injected instructions=========
6   asr r3 ; r3=r3/2
7   com r3 ; Take one's complement of r3
8   adc r3, r2
9   sbc r3, r2
10  ;==============
11  eor  r2, r3 ; Bitwise exclusive or
           between r2 and r3
12
13  ; the rest are same as "Program B"
14  ...
```

```
1   ...
2   ; up to here, same as "Program B"
3
4   add  r1, r2 ; Add r2 to r1 (r1=r1+r2)
5   ;===== 2 injected instructions=========
6   asr r3 ; r3=r3/2
7   com r3 ; Take one's complement of r3
8   ;
9   ;
10  ;==============
11  eor  r2, r3 ; Bitwise exclusive or
           between r2 and r3
12
13  ; the rest are same as "Program B"
14  ...
```

Figure 5. The version of *Program B* after the injection of 4 malicious instructions (left) and the same program after the injection of 2 malicious instructions (right). The latter is considered a harder case.

### 4.3 Dataset Structure

We captured 1000 instances of each one of the programs (four in total) using the setup mentioned in subsection 4.1. Due to the difference in the number of instructions in each program, the total number of samples varied among the captured sets. Details about the records in the datasets are included in Table 1.

According to the task, we synthetically generated instances of *Program B* i.e., the benign modified version of the program. The reader should note that we also captured real EM samples for *Program B*, to provide ground truth for our comparative evaluation. Examples of the captured signals, and their corresponding instructions, are illustrated in Figure 6. These sets of data were split into certain combinations for evaluation purposes and were subjected to pre-processing.

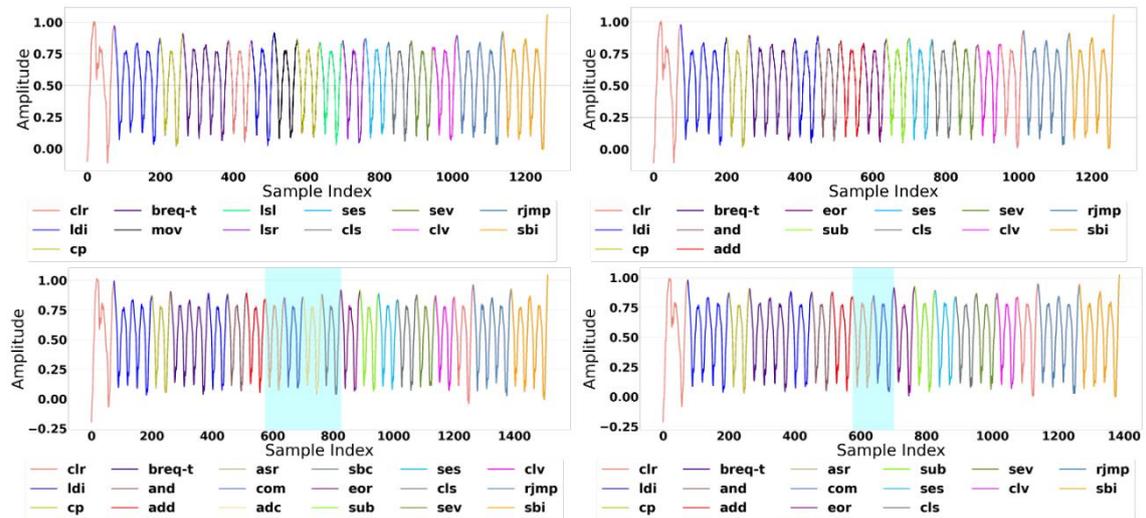

Figure 6. EM samples of the two normal (top row) and malicious (bottom row) programs. In the bottom row, the highlighted areas are the instructions injected into the base code of *Program B*.

### 4.4 Experimental Results

As a first experiment, we relied upon *only real examples of signals*. More specifically, the real signals that correspond to the normal programs (before and after the update) were used to train the baseline. Thus, at this phase, no malicious observations were used. During the testing phase, examples of both the normal and malicious cases were used. In fact, we performed two rounds, one in which the examples of malicious signals were drawn from the pool of signals that corresponds to the *easier* case. For the second round, the malicious signals were chosen from the pool of *harder-to-detect* anomalies. The goal of this experiment was to estimate the accuracy of the anomaly detection method in the ideal situation where real signals are available.

A second experiment was performed in a similar fashion to the first, except that *real examples for the original program* and *synthetic data for the modified version* were used to train the baseline. The goal of the second experiment is to approximate the predictive performance when utilizing synthetic signals, at least for one program/branch.

The two experiments were evaluated using the 10-fold cross-validation method. For experiment one, for each fold, the training set was comprised of 450 examples of *Program A* and 450 of *Program B*. Furthermore, to evaluate with a balanced testing set, the testing dataset considered for each fold contained only 50 observations of each benign (original and modified) case along with 100 anomalous examples. For the second round, the number of signals of each different type of programs used was the same for the training/testing set was the same except that the training set contained synthetic EMs for *Program B*.

**Pre-processing:** Before the training and testing phases, feature engineering was performed. First, every signal was reduced to the size of the benign execution paths. The reader should keep in mind that the size of the benign sequence is known in advance. As such, we assume that every signal that is being evaluated should only be the size of the benign case if it truly is benign. The reader should recall that each instruction is amplitude modulated. Therefore, the main indicator for identifying various instructions is the difference in the amplitude of the signal at certain time frames (i.e., cycles). In fact, one challenge that we observed in raw signals is that occasionally there are minor clock drifts. By maintaining only the peaks, we effectively deal with this issue without relying on computationally heavy techniques such as Dynamic Time Warping (DTW).

**Considered Parameters:** After performing a grid search, we identified the optimal *nearest-neighbors* parameter to be $\kappa = 10$. Moreover, the anomaly detection process made use of thresholds $\tau$ ranging from zero to one, with a step of $0.001$.

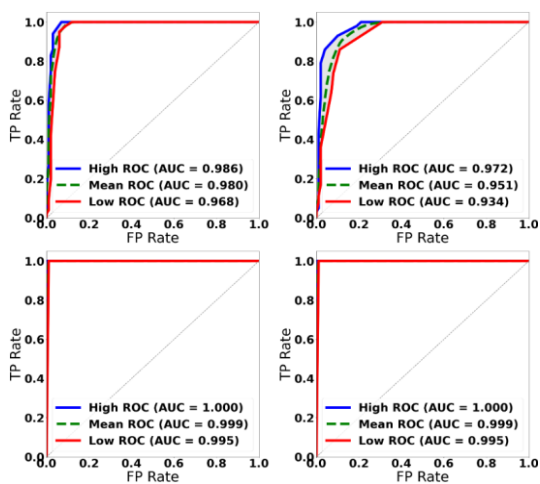

Figure 7. ROC graphs for the anomaly detection experiment. In the upper row, the results were obtained when using the synthetic signals for training. In the lower row, the results were obtained when using only real signals for training (ideal case). The drop in AUC score observed is only

1.3% for the injection of 4 instructions (the easy case) and 4.2% for the injection of 2 instructions (the hard case).

**Evaluation Metrics:** Given the confusion matrix results, we obtained the Area under the Curve (AUC) of the Receiver Operating Characteristic (ROC), and among the thresholds tested the one that gives the best accuracy (ACC) and F1 score for each fold was considered. The final reported metrics are the average among all folds. The max, minimum, and average ROC curves observed across all folds are given in Figure 7.

**Results:** The results achieved for each of the experiments can be seen in Table 2. Using *synthetic* data gives above 90% AUC score for all considered metrics. More specifically, the AUC score achieved when using the easy malicious case is 98%, and 95.1% when using the hard version. The AUC score achieved for the same tests when *real* signals were used is 99.3% for both the easy and hard cases. In other words, the use of synthetic signals had a negative impact on the predictive AUC, but it was relatively low i.e., 1.3% and 4.2% respectively. The reader should recall that despite the malicious programs being labeled *easy* and *hard* both cases correspond to exceptionally minimal injections and in reality, the attacker probably would try to inject much larger lengths of instructions. In terms of ACC and F1 scores, the use of synthetic signals achieved 90.1% and 90.6% respectively when using the hard malicious case. Furthermore, these metrics improve when evaluating against the easy version to 95.4% for the ACC and 95.5% for the F1 score. When the same tests are performed using the real signals, the ACC and F1 is near perfect, that is 99.9% and 99.5% respectively for the hard case and 99.8% for the easy version. Overall, the difference in the use of synthetic signals was 4.5% to 9.8% for the ACC and 4.3% to 8.9% for the F1.

Table 2. Results of the two experiments performed.

| Training | Test-Normal | Test-Anomaly | Scores (Avg.) | | |
|---|---|---|---|---|---|
| | | | AUC | ACC | F1 |
| Real A and Synthetic B | Real A and Real B | Malicious B (Hard) | 0.980 | 0.954 | 0.955 |
| | | Malicious B (Hard) | 0.951 | 0.901 | 0.906 |
| Real A and Real B | Real A and Real B | Malicious B (Hard) | 0.993 | 0.999 | 0.998 |
| | | Malicious B (Hard) | 0.993 | 0.999 | 0.995 |

**Conclusion:** *While using real captured EM samples may provide near-perfect detection of even minimal code injections, synthetic fingerprinting can still effectively train models to distinguish between benign and anomalous cases with high accuracy. For example, the penalty in terms of AUC score is -1.3% for detecting only four malicious instructions.*

## 5. DISCUSSION

One of the most important challenges with respect to the synthetic reconstruction of signals from code is that the morphological characteristics of isolated ASM instructions are not static but rather depend on prior instructions. Moreover, they possibly get influenced by other random events that occur at the hardware level or due to parallel processes that are executed at the same time (software), as well as environmental noise. Studying the first two factors lies outside the scope of this paper while the latter has been studied in [12], [15]. However, regarding the impact of previous instructions, we have made the following observations:

- Although the same instructions may have roughly the same amplitude and general phenotype when observed within the same sequence, they *may appear different* when preceded by different previous instructions or tracked within a different sequence.
- The directly previous instruction $I_{i-1}$ impacts the examined instruction $I_i$ *significantly* but in some cases even previous instructions $..., I_{i-2}$ may impact $I_i$ to a lesser extent.
- Certain instructions impact subsequent instructions *less* than others.
- Instructions that perform *similar operations* may impact subsequent instructions in a similar way.

To better understand the practical impact of the challenges outlined previously let us give some concrete examples as they were observed in our datasets.

**Example 1**: The sequence $\{..., ses, cls, ser, clv, ...\}$ is observed in both our benign programs. However, for *Program A* the instruction $ses$ is preceded by the $lsr$ instruction while in *Program B* the $ses$ instruction is preceded by the $sub$ instruction. Nevertheless, both the $lsr$ and $sub$ instructions perform similar (i.e., mathematical) operations. The former performs division and then shift, while the latter performs subtraction. Therefore, the amplitude of the first instruction in that sequence, (i.e., the $ses$ instruction) is only marginally impacted.

**Example 2**: The sequence $\{..., rjmp, sbi, ...\}$ is observed in both the considered benign programs. In this case, for *Program A* the $rjmp$ is preceded by the $clv$ instruction, while in *Program B* the same instruction is preceded by the $clr$ instruction. The former simply clears the value of a flag while the latter resets the values of all registers. The reader can understand that the two instructions perform drastically different operations thus, it does not come as a surprise that the amplitude of the signal that corresponds to the $rjmp$ instruction looks significantly different in the two programs. A comparison between the signals corresponding to the two programs at the sections of interest is given in Figure 8.

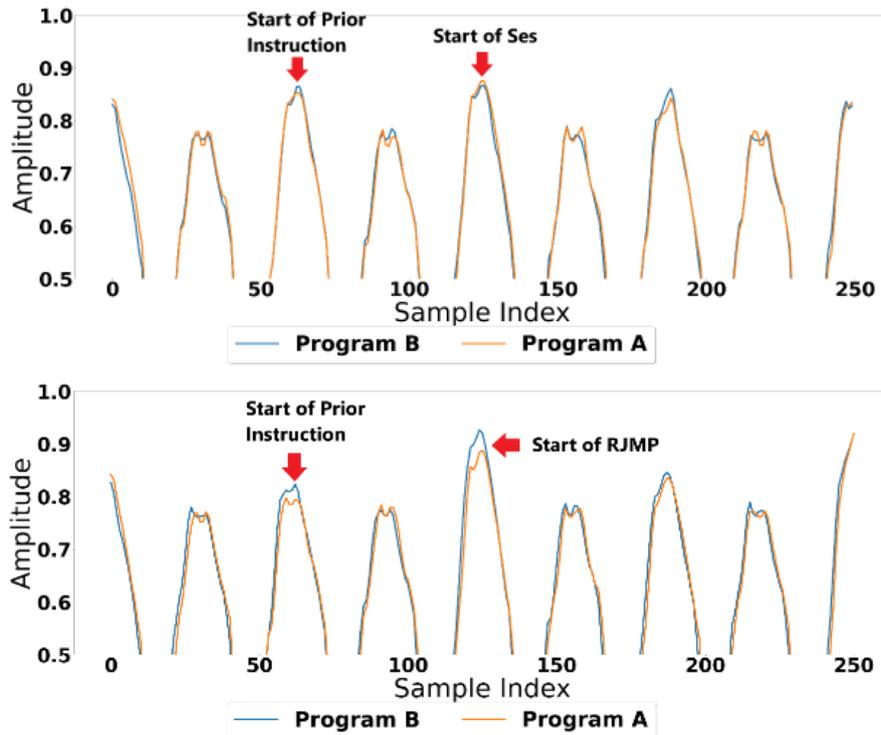

Figure 8. Comparison of peaks between EM signals of *Program A* and *Program B* around the common instruction sequences for example 1 (top) and example 2 (bottom).

## 5.2. Considerations about the Library of Reusable Basic Blocks

Let us suppose that the previous observations were not true. Then, it would be possible to construct a set of programs comprised of the instruction to be fingerprinted $I_i$ surrounded by sequences of $nop$ instructions as:

$$P_i = \{\dots, nop_{n-2}, nop_{n-1}, nop_n, I_i, nop_{n+1}, nop_{n+2}, \dots\}$$

Notice $nop$ instructions are considered neutral as they do not perform any function but simply consume a cycle thus, they are an ideal choice for this fingerprinting task. For the considered CPU architecture this would amount to creating 123 unique programs i.e., the same as the number of unique instructions. At a subsequent step, the instruction $I_i$ would be stripped from surrounding the $nop$ and entered in a database. In the future for any given sequence of instructions, it would be possible to consult this database and retrieve the corresponding EM sequences. In this scenario, the entire workflow is deemed trivial, and the task of EM synthesis is reduced merely to simple mapping.

However, as explained in the previous subsection, in reality, the task is not trivial because each instruction $I_i$ in a sequence is influenced at least by the previous instruction $I_{i-1}$. Thus, the database of reusable components must be constructed by considering at least two instructions. The situation because more challenging because in turn instruction $I_{i-1}$ is expected to have been altered by the effects of $I_{i-2}$.

To put things into perspective, for our considered CPU architecture the number of possible instruction combinations is $123x123$ which is more than two orders of magnitude larger than the naïve case. Alternative CPU architecture may support a significantly higher number of instructions. It is obvious that this approach does not scale. However, particularly for the embedded realm, creating a database of this type should *not* be considered prohibitive because (a) the majority of CPU architectures adopt a reduced set, (b) it is possible to identify similar instructions and cluster them, (c) in practice, not all combinations of certain instructions make sense.

## 5.3. Similarity of Synthetic & Real Signals

The following experiment evaluates the similarity of the 1000 EM signals that were synthetically generated against each type of each of the sets that contain real signals (i.e., *Program A, Program B, Malicious Easy, and Malicious Hard*). For each set, the comparison process yields 1000 similarity scores. The scores of the 25 nearest neighbors were averaged. The pre-processing method adopted, was the same as in all previous experiments. The distance used to compare the two signals was the Normalized Euclidean Distance (NED). This metric is calculated as:

$$NED(A, B) = \sqrt{0.5 \frac{Var(A - B)}{Var(A) + Var(B)}}$$

where *A* and *B* are two EM instances, and *Var* is the variance between the two signals. The distances produced as a result of this experiment are given in Figure 9. By observing the boxplot, the reader should notice that the average similarity distance (i.e., orange line) between *Program B* is lower compared to any other program. This indicates that our method generates signals that are closer to the real instances of *Program B* although clearly not identical to them. Furthermore, *Malicious Easy and Malicious Hard* cases have on average a much higher distance (difference) to *Synthetic B*, despite being polluted with only a few instructions. This is primarily because the injection of instructions causes a displacement to the right of all instructions after the point of injection. For this reason, all peaks after that point are expected to be different. On the other hand, the difference between Program A and Program B lies primarily in the (benign) substitution of some instructions. In this way, only the substituted instructions are expected to be different.

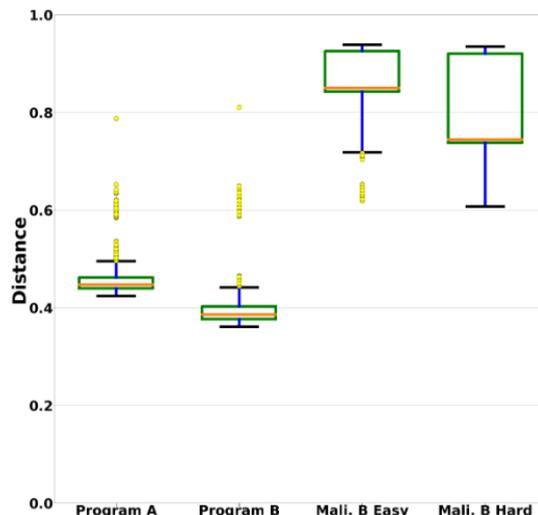

Figure 9. Distance between the synthetically generated version of Program B and (a) the real Program A, (b) the real Program B, (c) the maliciously modified version of Program B (4 instructions), and (d) the maliciously modified version of Program B (2 instructions). Lower is better (more similar).

## 6. Conclusions

In this paper, we introduced a comprehensive framework for generating synthetic EM signals from machine code and used it for EM-based anomaly detection in embedded devices. Compared to the state-of-the-art in the area, our approach remains non-intrusive and is highly scalable. We experimentally proved that our method can generate synthetic signals that are highly similar to the real EM signals that get emanated by the CPU of embedded devices during run-time. Our method inflicts only a small penalty in accuracy when employed for anomaly detection purposes. While the experiments included were performed based on a limited number of instructions, a limited number of test cases, and just one CPU architecture, the results achieved hold great promise for the utilization of synthetically generated signals as part of typical anomaly detection workflows in the area.

Our near-term research plans are geared towards quantifying the impact of various instructions on the morphology of subsequent ones. This study will help to create more accurate synthetic signals. In the future, to further automate the process and achieve even more accurate results we are considering incorporating Generative Adversarial Networks (GANs) specifically for the task of translating text (ASM code) to signal (EM).

## References

[1] S. D. D. Anton, A. P. Lohfink, and H. D. Schotten, "Discussing the Feasibility of Acoustic Sensors for Side Channel-aided Industrial Intrusion Detection: An Essay," in Proceedings of the Third Central European Cybersecurity Conference, 2019, pp. 1–4.

[2] D. Barbará, C. Domeniconi, and J. Rogers, "Detecting Outliers using Transduction and Statistical Testing," in Twelfth ACM SIGKDD International Conference on Knowledge Discovery and Data Mining, Philadelphia, 2006, pp. 55–64.

[3] H. A. Khan, M. Alam, A. Zajic, and M. Prvulovic, "Detailed tracking of program control flow using analog side-channel signals: a promise for iot malware detection and a threat for many cryptographic implementations," in Cyber Sensing 2018, 2018, vol. 10630, p. 1063005.

[4] P. Guide, "Intel® 64 and ia-32 architectures software developer's manual," Volume 3B: System programming Guide, Part, vol. 2, no. 11, 2011.

[5] Y. Han, S. Etigowni, H. Liu, S. Zonouz, and A. Petropulu, "Watch me, but don't touch me! contactless control flow monitoring via electromagnetic emanations," in Proceedings of the 2017 ACM SIGSAC conference on computer and communications security, 2017, pp. 1095–1108.

[6] C. Hunger, M. Kazdagli, A. Rawat, A. Dimakis, S. Vishwanath, and M. Tiwari, "Understanding contention-based channels and using them for defense," in 2015 IEEE 21st International Symposium on High Performance Computer Architecture (HPCA), 2015, pp. 639–650.

[7] C. Kolias, D. Barbará, C. Rieger, and J. Ulrich, "EM Fingerprints: Towards Identifying Unauthorized Hardware Substitutions in the Supply Chain Jungle," in 2020 IEEE Security and Privacy Workshops (SPW), 2020, pp. 144–151.

[8] C. Kolias, R. Borrelli, D. Barbara, and A. Stavrou, "Malware detection in critical infrastructures using the electromagnetic emissions of plcs," Transactions, vol. 121, no. 1, pp. 519–522, 2019.

[9] Y. Liu, L. Wei, Z. Zhou, K. Zhang, W. Xu, and Q. Xu, "On code execution tracking via power side-channel," in Proceedings of the 2016 ACM SIGSAC conference on computer and communications security, 2016, pp. 1019–1031.

[10] Microchip, "AVR Instruction Set Manual." Accessed: Sep. 30, 2022. [Online]. Available: http://ww1.microchip.com/downloads/en/devicedoc/atmel-0856-avr-instruction-set-manual.pdf

[11] Microchip, "Microcontrollers, Digital Signal Controllers and Microprocessors." Accessed: Sep. 30, 2022. [Online]. Available: https://www.microchip.com/en-us/products/microcontrollers-and-microprocessors

[12] E. Miller, "Detecting Code Injections in Noisy Environments through EM Signal Analysis and SVD Denoising," PhD Thesis, 2022. [Online]. Available: https://www.proquest.com/dissertations-theses/detecting-code-injections-noisy-environments/docview/2674042935/se-2

[13] A. Nazari, N. Sehatbakhsh, M. Alam, A. Zajic, and M. Prvulovic, "Eddie: Em-based detection of deviations in program execution," in Proceedings of the 44th Annual International Symposium on Computer Architecture, 2017, pp. 333–346.

[14] N. Sehatbakhsh, M. Alam, A. Nazari, A. Zajic, and M. Prvulovic, "Syndrome: Spectral analysis for anomaly detection on medical iot and embedded devices," in 2018 IEEE international symposium on hardware oriented security and trust (HOST), 2018, pp. 1–8.

[15] K. Vedros, G. M. Makrakis, C. Kolias, M. Xian, D. Barbara, and C. Rieger, "On the Limits of EM Based Detection of Control Logic Injection Attacks In Noisy Environments," in 2021 Resilience Week (RWS), 2021, pp. 1–9.

[16] S. Wei, A. Aysu, M. Orshansky, A. Gerstlauer, and M. Tiwari, "Using power-anomalies to counter evasive micro-architectural attacks in embedded systems," in 2019 IEEE International Symposium on Hardware Oriented Security and Trust (HOST), 2019, pp. 111–120.